\documentstyle[prl,preprint,epsf,aps]{revtex}
\tighten
\begin{document}
\title{Electronic Dephasing in the Quantum Hall
Regime}
\author{N. A. Fromer, C. Sch\"uller\cite{schul}, and D. S. Chemla}
\address{Department of Physics, University of
California at Berkeley, \\ and Materials Sciences
Division, Lawrence Berkeley National Laboratory,
Berkeley, CA 94720}
\author{T. V. Shahbazyan, and I. E. Perakis}
\address{Department of Physics and Astronomy,  
Vanderbilt University, Nashville, TN 37235}
\author{K. Maranowski and A. C. Gossard}
\address{Department of Electrical and Computer
Engineering, University of California at
Santa Barbara, Santa Barbara, CA 93106}

\date{\today}
\maketitle

\begin{abstract}
By means of degenerate four-wave mixing (FWM), we
investigate the quantum coherence of electron-hole
pairs in the presence of a two-dimensional electron
gas in modulation--doped GaAs-AlGaAs quantum wells
in the quantum Hall effect regime.
With increasing magnetic field, we observe a
crossover from Markovian to non--Markovian behavior, 
as well as large jumps in the decay time of the FWM
signal at even Landau level filling factors. The
main observations  can
be qualitatively reproduced by a model which takes
into account scattering by the collective excitations
of the two-dimensional electron gas.
\end{abstract}

\pacs{PACS~ 78.47.+p,78.66.Fd,73.20.Dx,78.20.Ls,42.50.Md}


The two-dimensional electron gas (2DEG) in modulation--doped
semiconductor quantum systems is a subject of great and still 
growing interest since it allows, in specially
tailored systems, the investigation of fundamental
properties of electrons in reduced dimensions. 
In particular, in a strong magnetic field, 2D electrons form a
strongly correlated system that exhibits 
such unique electronic transport properties as 
the integer and fractional quantum Hall effect.
Linear optical experiments were
successfully used to highlight these regimes of the 2DEG
\cite{kukushkin94}. However, very little is known
about the dynamics of such strongly correlated
electron systems.
On the other hand, ultrafast time resolved
nonlinear spectroscopy (TRNS) provides unique and
powerful tools for studying the dynamics of Coulomb
correlation effects in semiconductors. Because the
time resolution now reached is much shorter than the 
scattering times of elementary
excitations and the period of phonons or plasmons,
TRNS is well suited to investigate processes
that are much more difficult to access through
transport measurements, such as dephasing and
dissipation\cite{eisenstein92}. In fact, it was
demonstrated through TRNS that in undoped semiconductors
both coherent and dissipative processes are governed
by many--body effects \cite{chemla99}. Recently,
four--particle and higher order Coulomb correlation
effects, that could not be explained within the
time dependent Hartree--Fock approximation (HFA) have
been observed\cite{kner98}. Compared to the wealth of
experiments on intrinsic systems, rather few
investigations of coherent dynamics in
modulation--doped quantum wells, which
contain a cold 2DEG, have been reported 
\cite{knox88}.

In this paper we investigate the quantum coherence
of the interband polarization in the presence of
a 2DEG in a multiple modulation--doped 
GaAs-AlGaAs quantum well structure in the
regime of the quantum Hall effect. 
We observe very strong variation of the
interband dephasing time, $T_2$, as a function
of the filling factor, as well as direct evidence
of memory effects in the optical dynamics. 
In a strong magnetic field, such
that the 2DEG occupies only the lowest Landau level
(LLL), there are no interactions between the 
photoexcited pairs, unless there is an asymmetry
between electron and hole wavefunctions
\cite{lerner}. When the LLL is partially filled, the
dephasing originates mainly from the scattering of the
photoexcited carriers with the intra-LL collective
excitations of the strongly correlated 
2DE-liquid\cite{gmp,haus96,pinczuk93}. 
We present a model based 
on magnetorotons that accounts 
for most of the observed effects.

We first present the experimental results and
then discuss their interpretation. The samples
are multiple period, modulation--doped quantum wells,
antireflection coated and mounted on sapphire
substrates for transmission measurements. We
performed measurements on two samples
whose active regions have thirty periods, each
consisting of a 12 nm GaAs well and a 42 nm
AlGaAs barrier, where the center 12 nm is doped
with Si. The doped carrier density, $n$, under 
illumination is $n =2.6 \times 10^{11}$ cm$^{-2}$ 
in sample A, and $n=4.9 \times 10^{11}$ cm$^{-2}$
in sample B. Both samples had low temperature 
mobilities of $\mu \approx 8 \times 10^4$ cm$^{2}$/Vs. 
They were immersed in superfluid helium in an
optical split--coil cryostat at a temperature of
1.8 K. The four-wave mixing (FWM) \cite{chem2}
experiments were performed with two laser beams
of equal intensity which were in resonance with
transitions from valence--band to conduction--band
LL. We used spectrally narrow $\tau =300$ fs laser
pulses to resonantly excite only one LL in strong 
magnetic fields. The excitation intensity was kept 
low enough for the density of photogenerated $e$-$h$
pairs, $n_{eh}$, to remain small compared to
the doping density of electrons, typically 
$n_{eh} \lesssim n /10$. The beams were either 
left ($\sigma^-$) or right ($\sigma^+$) 
circularly polarized and separated
by a time delay $\Delta t$. At $B=0$, the linear
absorption spectra, $\alpha (\omega)$, exhibit
a clear Fermi-edge, at $E \approx 1.545$ eV in A
and at $E \approx 1.55$ eV in B. Only the LLL
is occupied ($\nu <2$), for $B> 5.1$ T in A and
for $B>9.8$ T for B. In time integrated (TI-FWM)
experiments, the total signal intensity was
measured by a PMT, while for spectrally resolved
experiments (SR-FWM) a spectrometer with 0.75 m
focal length and a CCD camera
were used to record the spectrum of the emitted
signal. 

Typical measurements of the TI-FWM signal,
$S_{TI}(\Delta t)$, in sample A are shown in Fig.\ 1,
for $B= 5.5$ T$ \rightarrow 11.5$ T using $\sigma^+$
polarized light (the $\sigma^-$ data shows the
same behavior). In these experiments the 
pulses were tuned to excite electrons
only into the highest partially occupied LL, which
contains the Fermi energy, $E_F$. For $5.5$ T$ \leq 
B \leq 6.5$ T the $S_{TI}(\Delta t)$ 
profile is a single exponential with an unusually 
long decay time. For $B \geq 7.5$ T the
profile is more complicated, showing non-exponential
behavior for short time delays. By extracting an overall
decay time we can get a direct measure 
of the interband polarization dephasing time $T_2$. 
The results are displayed
by the open circles in the upper panel of Fig. 2
for sample A and in the lower panel for sample B.
It is striking to note the very large jump of $T_2$
each time the system passes through even filling
factors and in particular at $\nu = 2$. Since these
features are reproducible as a function of
$\nu$ for samples with different densities, we
can assert that this is an effect of the
cold 2DEG. 

The non--exponential behavior of the TI-FWM signal
at high field is characterized by a change of slope
that occurs in sample A at
$\Delta t \approx 4.2$ ps$ \rightarrow 2.5$ ps as
$B \approx 7.5$ T$ \rightarrow 11.5$ T, indicating
memory effects in the polarization dynamics. 
These are also seen in the frequency domain
in Fig.\ 3, where we display the SR-FWM
signal, $S_{SR}(\Delta t,\omega)$ (together with 
$\alpha (\omega)$), at fixed $\Delta t=0$ for
$B= 5.5$ T$ \rightarrow 11.5$ T in Fig.\ 3(a), and at
fixed $B=11$ T for $\Delta t=0$ ps$ \rightarrow 6$ ps
in Fig.\ 3(b). Clearly the $S_{SR} (\omega)$ 
profile changes from a Lorentzian lineshape with 
a constant width,
$\Gamma \propto T_2^{-1}$, 
to an asymmetric one that would correspond
to a frequency dependent width, $\Gamma (\omega)$.
In Fig. 3(a) this occurs for $B \gtrsim 7.5$ T, and
in Fig. 3(b) for $\Delta t \lesssim 3$ ps.
Such a profile indicates a polarization relaxation term
$\propto \Gamma (\omega) P(\omega)$, which
gives in the time domain a dephasing
with memory structure, i.e.,
\begin{eqnarray} \label{scatt2}
\frac{\partial P}{\partial t}\Biggl|_{scatt} =
\, \int_{-\infty}^{t}dt' \Gamma (t-t') \, P(t').
\end{eqnarray}
We note also that if the $S_{SR} (\omega)$ spectra are
asymmetric, they are redshifted from the $\alpha (\omega)$,
while if they are Lorentzian they almost coincide
with the $\alpha (\omega)$ peaks.

The scattering rates for the density matrix elements,
${\hat \rho}$, i.e., interband polarization
and occupation numbers, can be calculated using the
general non--equilibrium formalism \cite{haug}. 
The memory kernel within the LLL can be presented as
$\Gamma (t-t') = (2\nu^{-1}-1) \kappa (t-t')$,
where the factor $(2\nu^{-1}-1)$, expected on general
physical grounds, is proportional to $N_s$, the
number of empty states available for scattering
within the LL containing $E_F$. It has the form
$N_{s} \propto (2(N+1) -\nu )/\nu$ in the $N$th LL
(factor of 2 for the spin). 
In addition to scattering with the intra-LL collective
excitations, there are several inter-LL relaxation
processes which contribute to the dephasing at weaker fields,
e.g., phonon and impurity scattering, Auger--like processes,
etc.\cite{auger}. These background processes lead
to Markovian dephasing, $\kappa(t)\rightarrow\delta(t)$,  
with $T_{2,bg}=[N_{s}F(B)]^{-1}$, where $F(B)$ depends
only weakly on $B$ (mainly via inhomogeneous LL
broadening).
We have plotted $N_{s}^{-1}$ in Fig.\ 2 (full circles),
normalized so that the maximum height coincides with
that of the $T_2$ curve.
For low fields the agreement is striking; however
there are significant differences in the
$B$-dependence of $T_2$ for strong field. In
particular, the change in behavior occurs for
sample A at $B\gtrsim 7.5$ T, where we begin
to see the non--exponential behavior in Fig.\ 1,
or the asymmetry in Fig.\ 3(a). Also, above this field,
the dephasing rate in Fig. 2 begins to differ
considerably from $N_s^{-1}$. 
Our analysis of the experimental data
to get the decay times of Fig.\ 2 is equivalent to
a Markovian approximation of Eq.\ (\ref{scatt2}).
We attribute this observed transition from Markovian
to non-Markovian behavior to a suppression of the
inter-LL scattering relative to
the dynamical response of the collective excitations
of the 2DE-liquid. 
At large magnetic fields, where the cyclotron energy,
$\hbar\omega_c$, is large compared
to other characteristic energies of the system, 
relaxation is dominated by intra-LL processes. Scattering
by collective excitations involves the matrix elements
of the dynamically screened interaction,
$U_{ij}^{<}(t,t')$, which in the LLL have the form:
\begin{eqnarray}\label{U}
U_{ij}^{<}(t,t')&=&
\int \frac{d{\bf q}}{(2\pi)^2}
e^{-q^2l^2/2}v_q^2\,\bar{\chi}_q^<(t,t')
c_{ij}(q) \, ,
\end{eqnarray}
where
$\bar{\chi}_q^<(t,t')=
\langle\bar{\rho}_{\bf q}(t')\bar{\rho}_{-\bf q}
(t)\rangle$ is the density--density correlation function
projected onto the LLL \cite{gmp,haus96}, and
$\bar{\rho}_{\bf q}(t)$ is the corresponding
density operator. Also, $v_q$ is the unscreened
Coulomb interaction, $l = (\hbar/eB)^{1/2}$ 
is the magnetic length, and the
$c_{ij}(q)$ with $i,j \rightarrow e,h$
model the asymmetry in the {\em e-e} and {\em e-h}
interaction matrix elements, which orginates from the
difference between electron and hole LLL wavefunctions.
Because of the breakdown of the perturbation
theory due to LL degeneracy in 2D systems at high fields, it
is incorrect to evaluate $\chi_q^<(t,t')$ 
within the
standard RPA \cite{haug}. Instead, one should account
for the true excitations of the interacting 2DE-liquid.
Several models can be found in the literature, and
we base our discussion on the magnetoroton model, 
which is the one best suited for our filling factors. The most
salient features are, however, general and model
independent. The magnetoroton dephasing mechanism
is somewhat similar to that of acoustic phonon
scattering. The details are presented
elsewhere \cite{tigr}, but we discuss here the general
trends. In our experimental conditions to a very good
approximation, the intra-LL collective excitations
are not affected by the small density of photogenerated
carriers, so one can use the equilibrium density
correlation function\cite{gmp}, and
\begin{eqnarray}\label{scatt1}
&&
\frac{\partial {\hat \rho}_{ij}}{\partial t}
\Biggl|_{scatt}=i\sum_{k}\int_{-\infty}^{t}dt'
G_{i}^{r}(t-t')G_{j}^{a}(t'-t)
\nonumber\\&&\times
\Biggl(
[U_{ik}^<(t-t') -U_{kj}^<(t-t')]
\rho_{ik}^<(t') \rho_{kj}^>(t')
- (<\leftrightarrow >) \Biggr),
\end{eqnarray}
where $G_{i}^{r/a}(t)$ is the retarded/advanced 
Green function,
$\rho_{ij}^< = \rho_{ij}$, 
and $\rho_{ij}^> = \delta_{ij}-\rho_{ij}$. 
If all $U_{ij}$ are equal, i.e., $c_{ij}(q)=1$,
then the polarization scattering term 
vanishes\cite{lerner}. This corresponds 
to identical
electron and hole wavefunctions in the LLL.
In practice, there is always asymmetry between
electrons and holes, due to, e.g.,  
differing band offsets, lateral confinement, and disorder. 
Using the results of Ref.\ \cite{gmp},
Eq.\ (\ref{U}) takes the form
\begin{eqnarray}\label{Umag}
U^{<}(t)=
-\frac{in}{2\pi}\int
&&
\frac{d{\bf q}}{(2\pi)^2}e^{-q^2l^2/2}v_q^2c_{ij}(q)
\nonumber\\&&\times
\bar{s}_q [(N_q+1)e^{i\omega_qt}+ N_qe^{-i\omega_qt}],
\end{eqnarray}
where $N_q$ is the Bose distribution function for 
magnetorotons of energy $\omega_q$, and $\bar{s}_q$ is 
the static stucture factor of the 2DE-liquid in the LLL.
By comparing Eqs.\ (\ref{scatt1}) and (\ref{scatt2}),
we see that the $\omega$ dependence of $\Gamma(\omega)$
is determined by the Fourier transform of $U^<(t)$, which
is governed by the $q$ dependence of $\bar{s}_q$. In the
LLL $\bar{s}_q =(2\nu^{-1} - 1)\tilde{s}_q$\cite{spinexp} where 
$\tilde{s}_q \sim (q l)^4$ for $ql\ll 1$,
$\sim \exp(-q^2l^2/2)$ for $ql\gg 1$, and  
$\tilde{s}_q$ displays a peak for $ql\sim 1$\cite{gmp} that
leads to the magnetoroton excitations. The corresponding
resonance in $\Gamma(\omega)$ near the magnetoroton
energy leads to non--Markovian behavior 
with a characteristic response time of approximately the inverse 
of this energy. The latter is
estimated from the gap at the magnetoroton dispersion minimum, 
$\Delta\sim 0.1(e^2/\epsilon l)$ for our range of $\nu$\cite{gmp}, 
which for $B=10$ T is $\approx 1.5$ meV.
The experimental data of Figs.\ 1 and 3 strongly support our
interpretation, since they imply a reaction time 
$T_r \approx 2.5$ ps$ \rightarrow 4$ ps for
the 2DE-liquid collective excitations. We note that this
corresponds to an energy $\approx 1$ meV$ \rightarrow 2$ meV.
Clearly, a much more involved
theoretical treatment is needed to identify the details
of the interaction processes in this regime\cite{tigr}.

The non-Markovian behavior of
2DEG excitations is well documented at zero field, where the ultrafast 
nonlinear response of a Fermi sea of electrons is determined
by the continuum of {\em e-h} pairs excited by the Coulomb potential 
of the photoinduced carriers. The small
characteristic energy of these excitations gives rise to a
non--adiabatic Fermi sea response leading to a non--exponential
polarization decay (absent in the HFA)\cite{ilias1}.
We also see here (in Fig.\ 3) similar effects 
in the $B$- and $\Delta t$-dependent shifts of the SR-FWM signal. 
For large field, e.g., $B=10.5$ T, $S_{SR}(\omega)$ is 
redshifted from the $\alpha(\omega)$ resonance due to 
a lowering of the 2DE-liquid energy by the attractive 
potential of a photoexcited hole, a process similar to that 
known for the Fermi edge singularity \cite{ilias2}. 
This dynamical redshift comes from the real part of the 
magnetoroton-induced self energy. Since the latter is also 
proportional to $N_s$, the redshift is absent for nearly filled LLL, 
i.e., at $\nu \approx 2$ or $B \approx 5.5$ T (in sample A);
the reason is that a 2DE-liquid in an incompressible state
cannot readjust to screen the hole potential.

In conclusion, we have 
investigated the quantum coherence of electron--hole pairs
in multiple period, modulation--doped GaAs-AlGaAs
quantum wells in the quantum Hall
regime. We observe a clear transition from Markovian to
non--Markovian behavior with increasing magnetic field. In the
former case, the dephasing is dominated by inter-LL electron
relaxation, and the $B$-dependence of the dephasing time follows that
of the number of available scattering states,  exhibiting peaks at
even Landau level filling factors. At high magnetic field, the 
FWM signal shows strong evidence of memory effects. We proposed 
a model based on scattering of the photoexcited electrons
with magnetoroton excitations in the lowest Landau level
that qualitatively accounts for the main features of
the experimental observations.


The authors are grateful to Lu Sham and Ming Wei Wu for
very helpful discussions. This work was supported by
the Alexander von Humboldt-Stiftung (C.S.) and by
the Director, Office of Energy Research,  Office of
Basic Energy Sciences, Division of Material Sciences
of the U.S. Department of Energy, under contract No.
DE-AC03-76SF00098.
T.V.S. and I.E.P. were supported by the NSF grant 
No. ECS--9703453 and by Hitachi, Ltd.




\newpage

\begin{figure}
\caption{\label{TI-FWM-fig}
TI-FWM signal measured in sample A
using $\sigma^+$ polarized light for
$B= 5.5$ T$ \rightarrow 11.5$ T. 
}
\end{figure}

\begin{figure}
\caption{\label{T2-B}
TI-FWM decay times versus magnetic
field for samples A and B. The open circles are
the experimental data points, and the filled circles
correspond to $N_s^{-1} = \nu/(2(N+1) - \nu )$, where $N$ is
the LL number and $\nu $ the filling factor. The data
was taken using  $\sigma^+$ polarized light.
}
\end{figure}

\begin{figure}
\caption{\label{SR-FWM-fig}
Gray shaded curves: SR-FWM signal, (a)
at fixed $\Delta t=0$ for
$B= 5.5$ T$ \rightarrow 11.5$ T, and (b)
at $B=11$ T for $\Delta t=0$ ps$ \rightarrow 6$ ps. The thick, unshaded
lines in (a) show the linear absorption spectra, $\alpha (\omega, B)$, for 
$B = 6.5$, $8.5$ and $10.5$ T.}
\end{figure}
\clearpage
\epsfxsize=5.0in
\epsffile{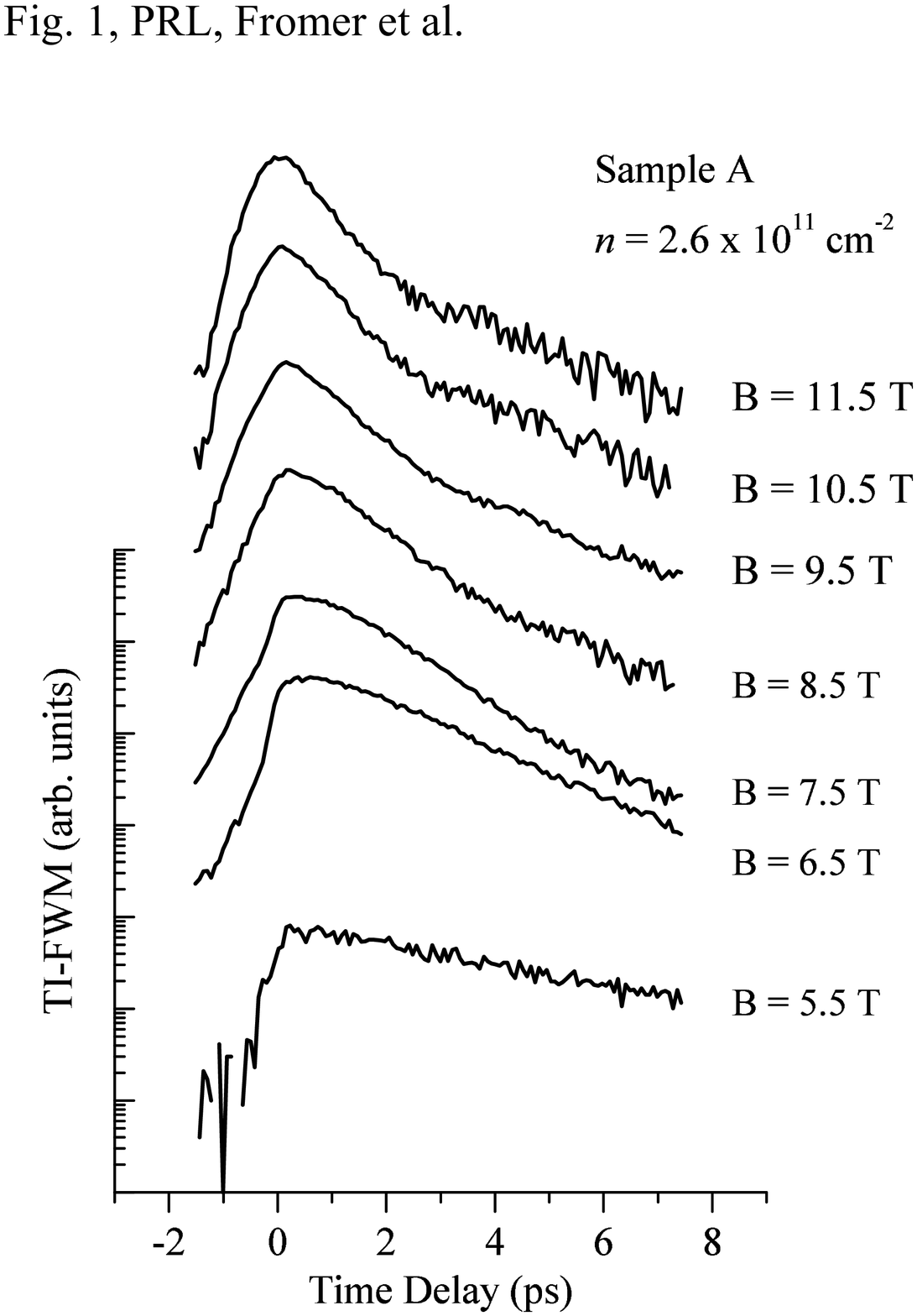}
\vspace{20mm}
\centerline{FIG. 1}
\epsfxsize=5.0in
\epsffile{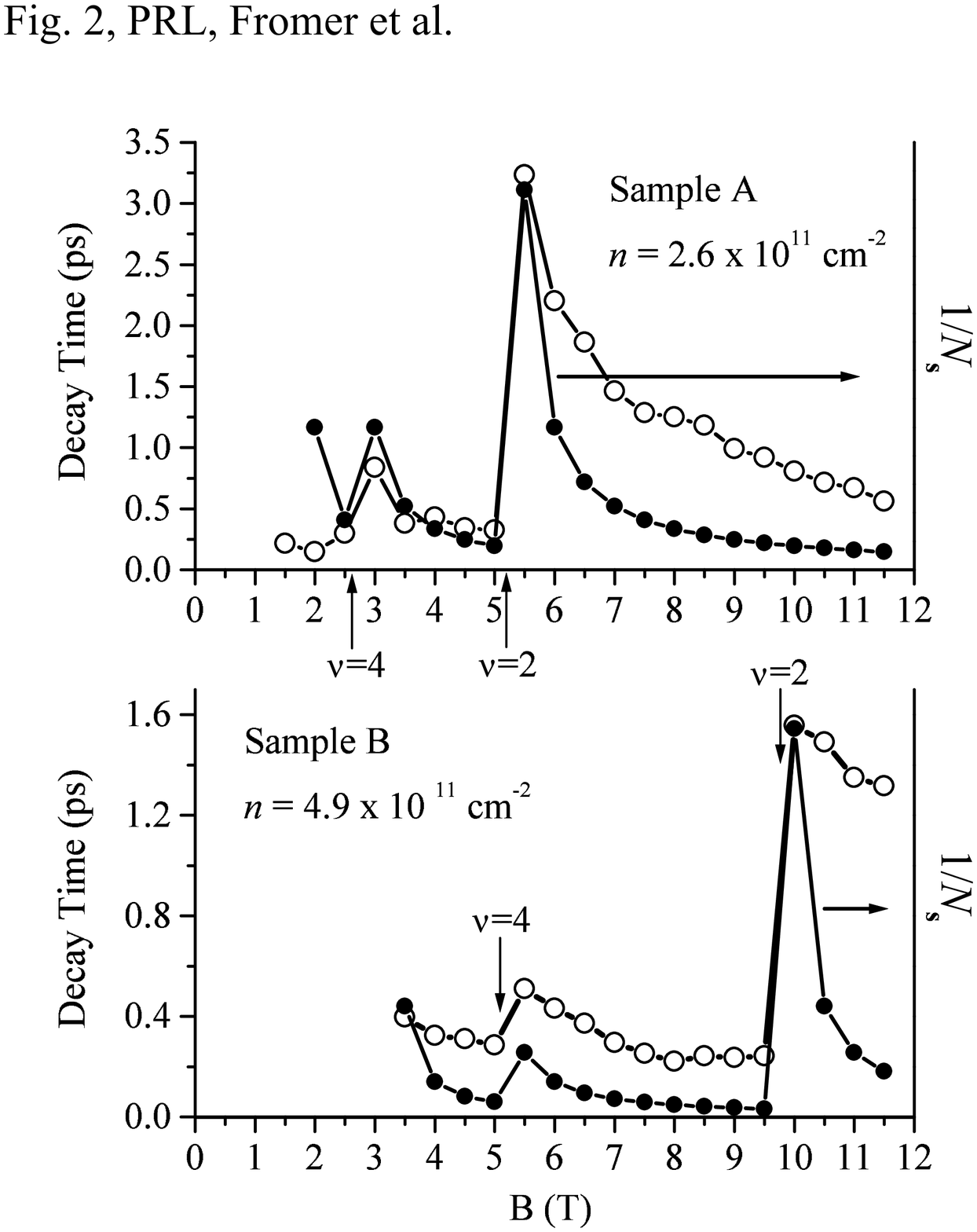}
\vspace{20mm}
\centerline{FIG. 2}
\epsfxsize=5.0in
\epsffile{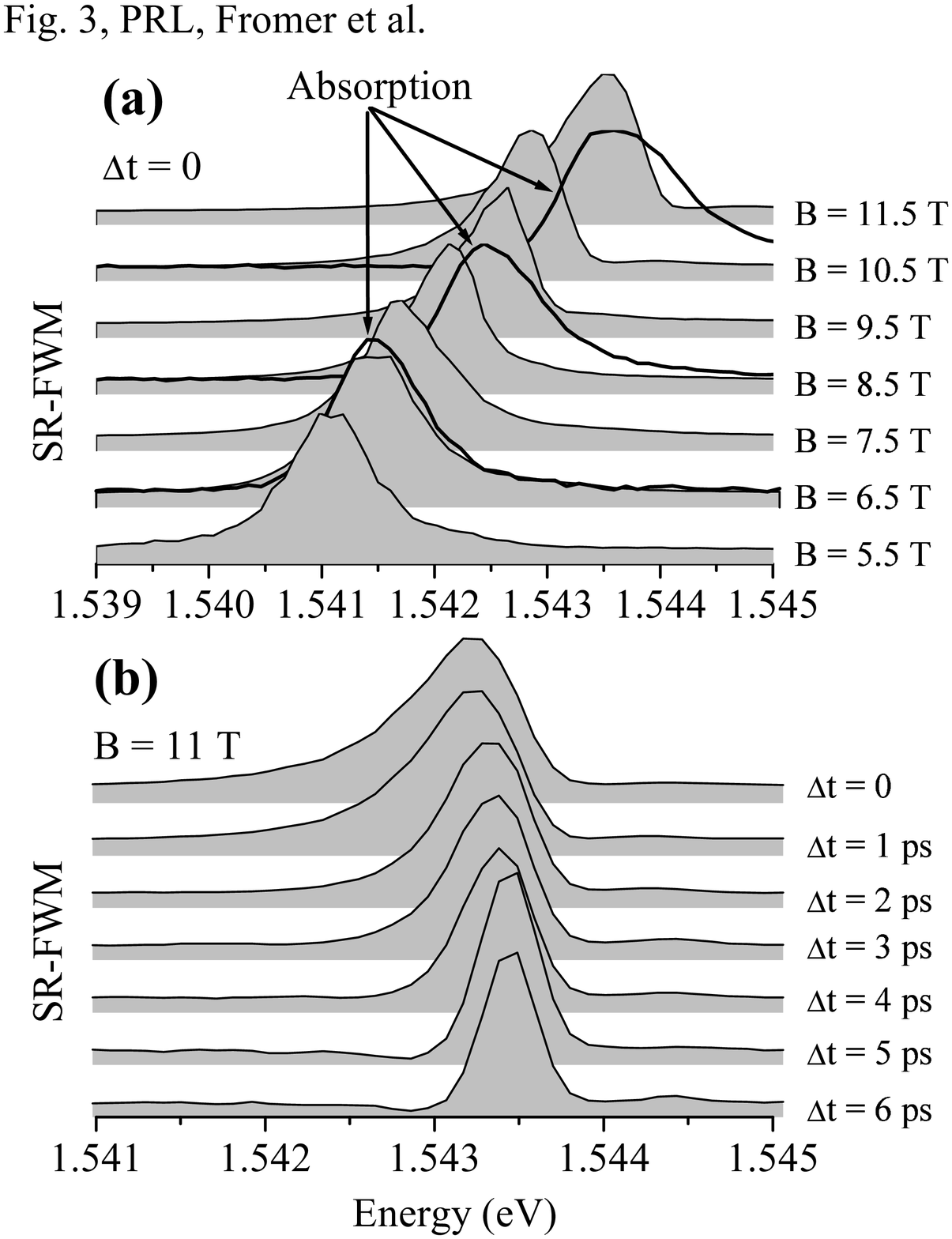}
\vspace{20mm}
\centerline{FIG. 3}


\begin{references}
\bibitem{spinexp} Here we follow Ref. [7], but
we include a factor of 2 since we neglect spin
splitting in the calculations.
\end{references}
\end{document}